\documentclass[conference]{IEEEtran}
\IEEEoverridecommandlockouts
\usepackage{cite}
\usepackage{amsmath,amssymb,amsfonts}
\usepackage{algorithmic}
\usepackage{graphicx}
\usepackage{textcomp}
\usepackage{xcolor}
\usepackage{hyperref}
\usepackage{gensymb}
\usepackage{subcaption}
\usepackage{adjustbox}
\usepackage{booktabs}

\def\BibTeX{{\rm B\kern-.05em{\sc i\kern-.025em b}\kern-.08em
    T\kern-.1667em\lower.7ex\hbox{E}\kern-.125emX}}

\begin{document}

\title{Overview of Deep Learning Methods for Retinal Vessel Segmentation\\
\thanks{}
}

\author{\IEEEauthorblockN{1\textsuperscript{st} Gorana Gojić}
\IEEEauthorblockA{\textit{The Institute for Artificial Intelligence } \\
\textit{Research and Development of Serbia }\\
Novi Sad, Serbia \\
gorana.gojic@ivi.ac.rs}
\and
\IEEEauthorblockN{2\textsuperscript{nd} Ognjen Kundačina}
\IEEEauthorblockA{\textit{The Institute for Artificial Intelligence } \\
\textit{Research and Development of Serbia }\\
Novi Sad, Serbia \\
ognjen.kundacina@ivi.ac.rs}
\and
\IEEEauthorblockN{3\textsuperscript{rd} Dragiša Mišković}
\IEEEauthorblockA{\textit{The Institute for Artificial Intelligence } \\
\textit{Research and Development of Serbia }\\
Novi Sad, Serbia \\
dragisa.miskovic@ivi.ac.rs}
\and
\IEEEauthorblockN{4\textsuperscript{th} Dinu Dragan}
\IEEEauthorblockA{\textit{University of Novi Sad}\\
\textit{The Faculty of Technical Sciences}\\
Novi Sad, Serbia \\
dinud@uns.ac.rs}
}

\maketitle

\begin{abstract}
Methods for automated retinal vessel segmentation play an important role in the treatment and diagnosis of many eye and systemic diseases. With the fast development of deep learning methods, more and more retinal vessel segmentation methods are implemented as deep neural networks. In this paper, we provide a brief review of recent deep learning methods from highly influential journals and conferences. The review objectives are: (1) to assess the design characteristics of the latest methods, (2) to report and analyze quantitative values of performance evaluation metrics, and (3) to analyze the advantages and disadvantages of the recent solutions.
\end{abstract}

\begin{IEEEkeywords}
medical imaging, retinal vessels, segmentation, deep learning, machine learning, overview
\end{IEEEkeywords}

\section{Introduction}

\begin{figure*}	
\includegraphics[width=\linewidth]{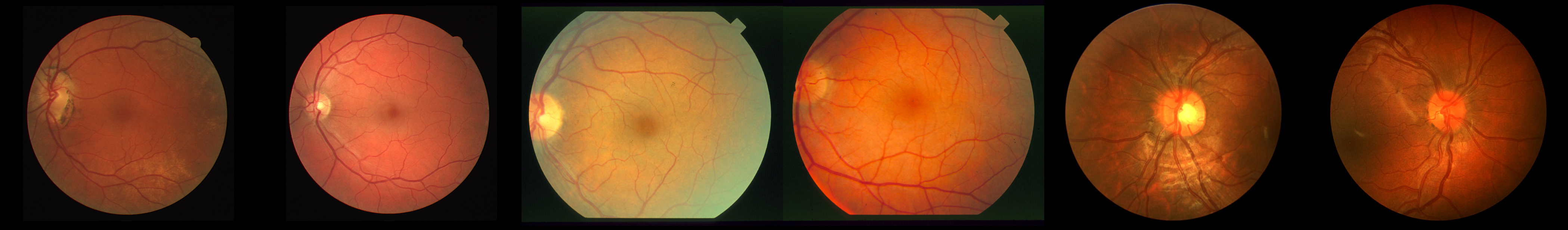}
\caption{Fundus images from DRIVE (images 1, 2), STARE (images 3, 4), and CHASE\_DB1 (images 5, 6) open fundus datasets.}\label{fig:fundus-image-samples}
\end{figure*}

Retinal vessel segmentation is a key step in the screening and early diagnosis of many eye and systemic diseases. Abnormal changes in the retinal vascular network may be indicators of retinopathy of prematurity \cite{cheung_2011_rop, gojic_2020_rop}, glaucoma \cite{fraz_2012_glaucoma}, age-macular degradation \cite{fraz_2012_glaucoma}, and diabetic retinopathy \cite{you_2012_diabetic_retinopathy}. 
Therefore, accurate segmentation of retinal vascularization is of great importance in screening and treatment procedures. Many imaging modalities can be used to capture retinal images, including fundus imaging (see Figure \ref{fig:fundus-image-samples}). A fundus image is a 2D projection of a 3D inner eye surface, with the retina being the top surface layer. Due to retinal semi-transparence property, the fundus image also shows non-informative anatomical structures belonging to layers below the retina often alleviating the segmentation process.

Some of the challenges in retinal blood vessel segmentation from fundus images are (1) low contrast, (2) image artifacts such as noise, blur, and uneven illumination, (3) variable blood vessel width and shape, and (4) blood vessel bifurcations and crossover points \cite{mookiah_2021_review, srinidhi_2017_review, petrovic_2022_robustness}. The aging of the world population and the increasing trend of vision impairment has resulted in an increased workload for ophthalmologists \cite{taylor_2001_world_blindness}. Increased workload leads to a higher probability of human error and increased risk to patient health. This has motivated higher research interest in automated retinal vessel segmentation methods to facilitate ophthalmologists in decision-making through computer-aided diagnosis (CAD) systems \cite{Hallak_2020_ai_revolution}. 
With the appearance of machine and deep learning methods novel approaches to retinal vessel segmentation capable of learning latent features have emerged. Recent years have demonstrated a growing development trend of these deep learning methods for retinal vessel segmentation \cite{mookiah_2021_review}. Combined with natural robustness improvement techniques \cite{gojic_2023_robustness}, these methods could enter broader clinical practice.

 
Retinal vessel segmentation methods are constantly evolving with new advancements being proposed frequently, especially in the deep learning domain. While review papers in the field already exist \cite{khandouzi_2022_review, mookiah_2021_review, chen_2021_review, soomro_2019_review, srinidhi_2017_review} they provide exhaustive reviews involving many studies from a variety of journals and conferences with varying degrees of reliability and reproducibility of reported results. A new review paper includes just the most recent publications from highly-influential conferences and journals providing insights into the latest development and trends relying solely on highly relevant studies in the field. This can help identify research gaps more efficiently and direct future research. Additionally, the review can serve as a quick introduction to the latest approaches to retinal vessel segmentation for researchers entering the field, but it can also serve as an extension to some of the existing review studies.



The objectives of the review are:
\begin{itemize}
    \item Summarizing research contributions of the most recent deep learning studies and discussing approaches used to address retinal vessel segmentation issues. 
    \item Summarizing and analyzing quantitative results reported in these studies in terms of retinal vessel segmentation evaluation metrics.
    \item Discussing advantages and disadvantages of the latest approaches, as well as further research directions.
\end{itemize}

The rest of the paper is organized as follows. In Section \ref{sec:rvs-theory} we give a definition of the retinal vessel segmentation task. Section \ref{sec:architecture-overview} gives an overview of DNN types that are commonly used in segmentation for this specific task. Follows the Section \ref{sec:numerical-results} with numerical results for commonly reported metrics in 39 papers. We provide a brief discussion in Section \ref{sec:discussion}, and conclude in Section \ref{sec:conclusions}.

\section{Retinal Vessel Segmentation}
\label{sec:rvs-theory}

\begin{figure}[!ht]
    \centering
    \includegraphics[width=\linewidth]{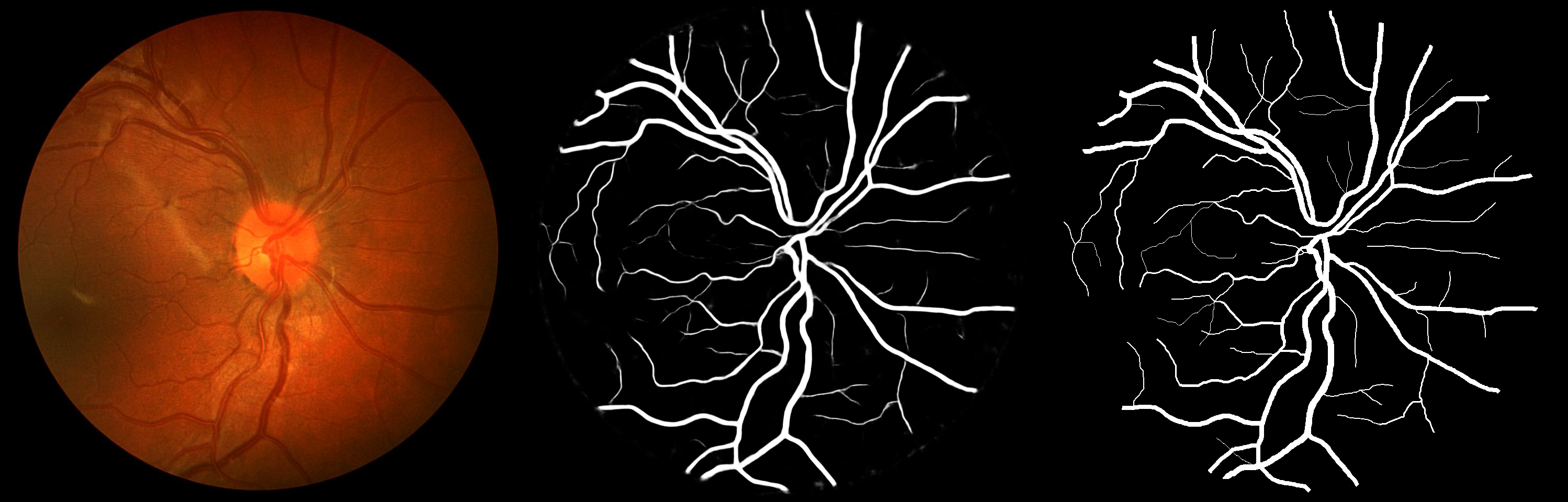}
    \caption{From left to right: a fundus image, a corresponding probability map, and a segmentation mask.}
    \label{fig:fundus-probmap-segmask}
\end{figure}

Retinal vessel segmentation from a fundus image is semantic segmentation in two classes, with \textit{blood vessel} and \textit{background} labels corresponding to the classes. Figure \ref{fig:fundus-probmap-segmask} shows an example of a fundus image, a corresponding probability map, and a segmentation mask. CNN outputs a probability map as an intermediate step in retinal vessel segmentation. The probability map is a grayscale image, with the color intensity of each pixel depicting the probability of the corresponding fundus image pixel being a blood vessel. A segmentation mask is obtained by binarizing the probability map. Binarization is performed by thresholding, or classifying all probability map pixels in one of two classes relative to the threshold. The threshold is either predefined or algorithmically determined separately for each probability map, e.g. using the Otsu algorithm \cite{otsu_1979_threshold}. For a specified threshold value, the segmentation mask is formed by assigning the \textit{blood vessel} label to a segmentation mask pixel if the color intensity of the corresponding probability map pixel is greater than or equal to the threshold, and the \textit{background} label if the color intensity of the pixel is smaller than the threshold.

\section{Overview of Segmentation Architectures}
\label{sec:architecture-overview}

In this section, we discuss different architectural choices for retinal vessel segmentation. We first introduce the UNet network which is a commonly used base architecture. Usually, it is modified to improve performance on the retinal vessel segmentation task by substituting existing layers with more efficient alternatives (e.g., dilated convolution) \cite{wu_2021_scsnet, wang_2020_rvsegnet} or by adding additional blocks (e.g., attention blocks) \cite{guo_2021_saunet, li_2021_fanet, zhang_2019_agnet, wang_2019_deunet}. We are focusing on a specific subset of CNNs for retinal vessel segmentation. These CNNs use attention mechanisms, context information, and domain knowledge to improve segmentation. We chose these techniques because they are among the most commonly used modifications to improve UNet segmentation results.

\subsection{UNet}
\begin{figure}
    \centering
    \includegraphics[width=\linewidth]{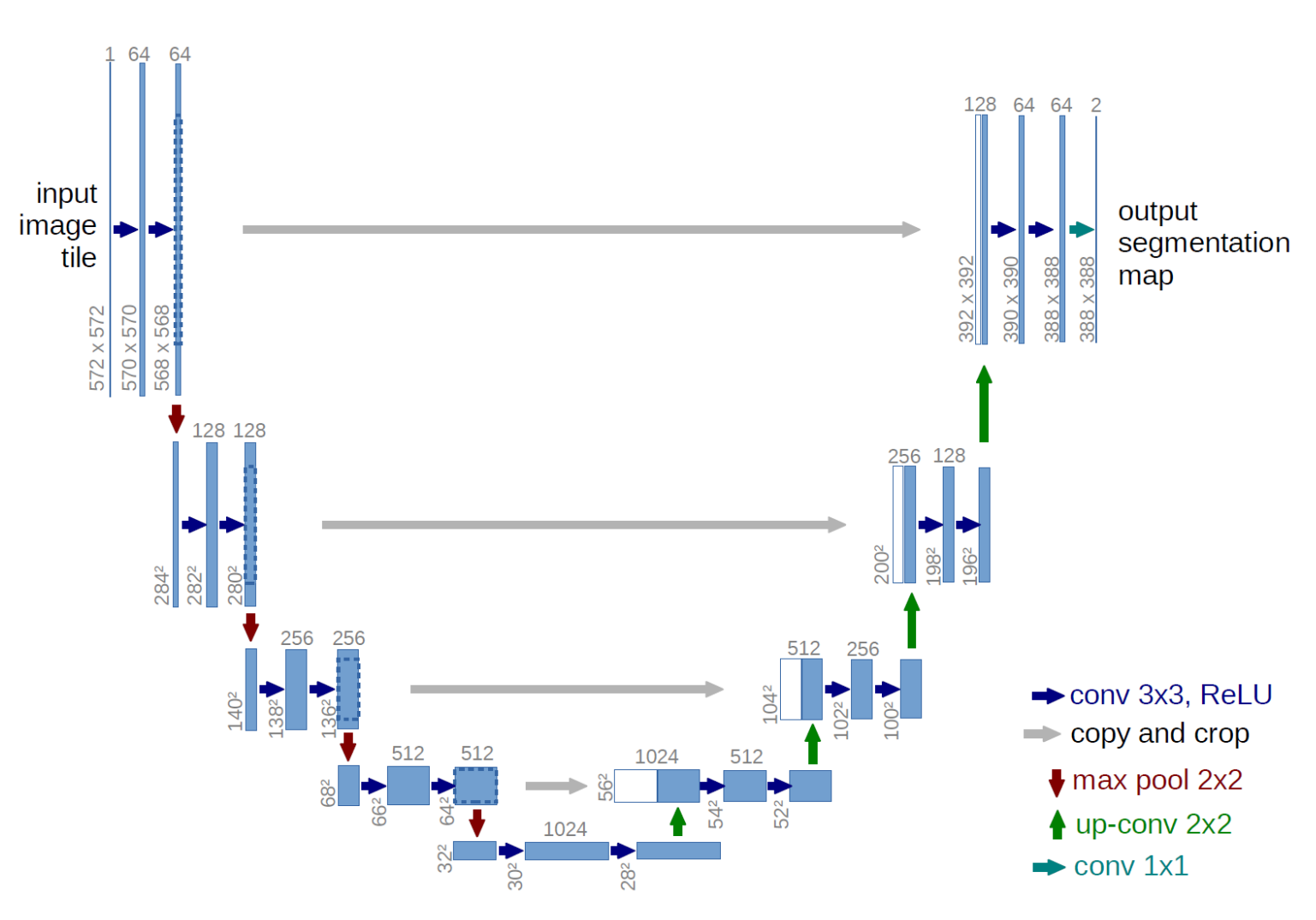}
    \caption{UNet network architecture. Illustration from \cite{ronneberger_2015_unet}.}
    \label{fig:unet}
\end{figure}

UNet architecture has been predominantly used in medical image segmentation. It is designed to accommodate small-size medical image datasets and strongly relies on data augmentation to compensate for the limited size of medical image datasets. UNet is a fully convolutional neural network with the encoder-decoder architecture shown in Figure \ref{fig:unet}. Here, the encoder learns to transform the input image into an embedding that is then expanded by the decoder into a probability map. The encoder of the originally proposed architecture in \cite{ronneberger_2015_unet} consists of five blocks, each performing convolutions and max-pooling operations to learn features of different complexity and reduce feature maps dimensionality. Each of the encoder blocks is directly connected to a corresponding decoder block by skip connections. Skip connections facilitate learning by providing higher-dimension encoder feature maps alongside lower-dimension decoder feature maps to generate decoder block output. In the originally proposed decoder blocks, max-pooling from the encoder block is replaced with up-convolutions to expand learn features into a probability map. 

The originally proposed architecture, like the most of the succeeding work in a field, predicts probability maps on a pixel level. Thus, pixel-wise loss functions like binary cross-entropy and dice loss are mostly used in training. Loss functions are commonly adapted to specific application needs like in \cite{ronneberger_2015_unet} where binary cross-entropy is modified to penalize errors in boundary pixel detection more severely. To address the class imbalance in retinal vessel segmentation, the loss function can be tailored to assign varying weights to errors in blood vessel and background pixel classification.


The majority of architectures for retinal vessel segmentation modify UNet to improve its performance. Some of the improvements include adding attention mechanism \cite{guo_2021_saunet, li_2021_fanet, zhang_2019_agnet, wang_2019_deunet} to weight the importance of learned features, exploiting knowledge about blood vessel appearance and geometry \cite{cherukuri_2020_geometry, yeon_2019_vgn, kipf_2017_semisupervised}, and using richer context information \cite{wu_2021_scsnet, wang_2020_rvsegnet, wang_2019_contextaware}.


\subsection{Domain-knowledge-based approaches}
The idea of incorporating domain knowledge in retinal vessel segmentation is motivated by the fact that learning CNNs in an unconstrained manner may result in learning features that are irrelevant to the segmentation task, e.g. the network might learn to extract different types of noise like morphological structures such as hemorrhages and microaneurysms. Domain-knowledge-based approaches offer a degree of performance robustness compared to approaches based solely on local image information. The performance of the latter is easily affected by the quality and quantity of training data that are critical in medical image applications. An additional challenge in retinal vessel segmentation is obtaining groundtruth masks for supervised learning, which remains the predominant approach in the field. Incorporating domain knowledge into segmentation architectures is an alternative to the attention mechanism since both approaches rely on learning a subset of features useful for segmentation. However, the former relies on information on geometry or pixel intensity, while the latter is completely domain-knowledge oblivious. 

One approach to exploiting domain knowledge is by introducing retinal vessel geometry structure in segmentation architectures \cite{cherukuri_2020_geometry,yeon_2019_vgn}. A work in \cite{cherukuri_2020_geometry} proposes a geometric representation layer that incorporates domain-specific knowledge to learn curvilinear features from fundus images. Knowledge is incorporated by modifying the loss function to include two regularizers to enhance learned kernel response on diverse orientations in the range 0-180\degree, and to penalize the generation of false positives. Additionally, the authors propose a policy to train the network to be robust to domain noise, such as hemorrhages and micro-aneurysms, by training on image patches that have a high chance of being interpreted as false positives. In \cite{yeon_2019_vgn} graph convolutional network (GCN) \cite{kipf_2017_semisupervised} is used to produce an additional set of features that are used jointly with local CNN features to generate final predictions. Here, a pretrained CNN is used to obtain a probability map that is used to generate an input graph for GCN by thresholding a probability map, applying skeletonization by morphological thinning, generating graph vertices by sampling over blood vessel pixels and generating edges between vertices. While having the advantage of providing vessel structural information, this approach is highly dependent on the quality of CNN-generated probability maps. It also has limited practical applicability, due to higher time and memory demands to run the segmentation pipeline.

Historically, the first to appear domain-knowledge approaches were hand-crafted filters \cite{sheng_2019_minspanningsuperpixel, wang_2017_colorfusion, zhang_2016_adaptivederivative, yin_2015_orientationawaredetector}. Here, we review just a single study on image matting that has been published recently. Image matting aims to extract a foreground given an image trimap, which is usually a manually annotated image having all pixels sorted out in the foreground, background, and unknown classes. Matting is formulated as a linear aggregation of a foreground and background image with coefficients $k$ and $1-k$ respectively, and the output of the algorithm are coefficient values \cite{fan_2019_imagematting}. Since it is demanding to create trimaps for fundus image vascularization, work in \cite{fan_2019_imagematting} proposes generating coarse trimaps automatically by preprocessing the input image and using coarse segmentation as an input to the image matting algorithm.

\subsection{Context-based approaches}

Context-based approaches rely on capturing more context information that would help generate more accurate segmentations. Here we discuss two implementations identified from the literature. The first includes strategies to adapt CNN's receptive field, as it has been shown previously that a fixed-size receptive field is not an optimal choice for multi-scale curvilinear structures as blood vessels \cite{wu_2021_scsnet, wang_2019_contextaware}. The second uses multi-scale feature extraction architectures to extract vessels of different diameters. Additionally, two implementations are often used jointly for optimal performance \cite{wu_2021_scsnet, wang_2020_rvsegnet, wang_2019_contextaware}.

To address adaptive receptive field size, \cite{wu_2021_scsnet} dynamically selects the most appropriate receptive field based on the characteristic of the target feature map. Paper contribution is a receptive field size selection criteria based on a correlation between learned weights for every two adjacent scales of the receptive field. However, the sole idea of choosing between multiple-sized kernels has been proposed earlier in \cite{li_2019_selective}, showing that CNNs incorporating selective kernels perform better in capturing multi-scale objects from natural images. Dilated, or atrous convolutions \cite{yu_2015_multi}, are used to capture more context by using a larger receptive field while keeping the same number of parameters compared to a regular convolution kernel. Dilated convolution, unlike regular convolution, introduces gaps in the receptive field. The extent of these gaps is determined by the dilation rate. When the dilation rate is set to one, dilated and regular convolution are equivalent. It has been used in multi-scale feature extraction in multiple papers on retinal vessel segmentation \cite{wu_2021_scsnet, wang_2020_rvsegnet}. In \cite{wang_2019_contextaware} optimal choice of context is learned through reinforcement learning \cite{goodfellow_2016_dl}. The proposed approach iteratively improves a given coarse segmentation mask by learning a policy to optimally sample image patches from the input image that are then segmented to assemble a segmentation mask. The policy determines patch extraction parameters, such as patch position and scale. By learning patch extraction properties instead using fixed patch extraction strategy, the approach shares the idea with \cite{wu_2021_scsnet} that learns receptive field size, but on a higher, image patch level.

The first layers of CNN learn high-resolution, detailed features based on the local context covered by receptive field size. Successive layers learn coarser, semantically meaningful features by aggregating the features learn from previous layers. In retinal vessel segmentation, CNN design results in reduced performance in thin blood vessel segmentation. To address this design choice, multi-scale implementations are proposed to handle blood vessel shape and diameter variability. Additionally, it is shown that some multi-scale architectures exhibit improved natural robustness on certain image corruptions and perturbations in natural images \cite{ke_2017_multigrid, huang_2018_multiscale}, which proves usable in segmentation from fundus images having different kinds of morphological noise structures \cite{wu_2021_scsnet}. Implementations of multi-scale architectures in retinal vessel segmentation vary. In \cite{wu_2021_scsnet} each encoder layer produces three output feature maps for a given input feature map, using dilated convolution with different atrous rates to cover different context areas. The resulting maps are concatenated into a single feature map passed to the next CNN layer. The multi-level semantic supervision module reconstructs probability maps from each decoder layer, using multi-scale feature maps to generate a more accurate segmentation mask. The authors also argue that using vanilla skip connections passes many non-informative, or even harmful features that the encoder learns, like noise. Thus, they propose an additional block to process features transferred by skip connections prior to decoding. Work in \cite{wang_2020_rvsegnet} proposes a feature pyramid cascade module to process encoder feature maps. Multi-scale average pooling is performed to obtain three feature maps, that are further processed by successive dilated convolutions and concatenated into an output feature map. A disadvantage of the multi-scale approach is increased computational and memory complexity, which is often alleviated through dilated or octave convolutions reducing parameter number compared to regular convolutions. Another approach is to limit the use of multi-scale features, such as in \cite{wang_2020_rvsegnet} where just the encoder output features are processed in a multi-scale manner.

\subsection{Attention-based approaches}
The attention mechanism in deep learning is based on the assumption that not all parts of the input data are equally important to solve the target task. The concept is originally introduced in language translation to eliminate a bottleneck problem, leading to performance loss for long input sequences. To alleviate the issue, parts of the input sequence are provided directly to the decoder when generating the output. Corresponding attention weights are learned for each part of the input, telling the decoder how important it is. Attention can be global if the entire input sequence is used to calculate the attention, or local if an input subsequence is used in the calculation. While specific implementations of the attention mechanism can vary, one of the most influential works on attention in sequence-to-sequence processing is introduced in Transformer architecture that uses attention solely to generate predictions \cite{vaswani_2017_attention}. The proposed attention mechanism has been successfully adapted to the computer vision field through vision transformers (ViT) \cite{dosovitskiy_2021_vit}. However, as ViTs have emerged recently, the majority of works still propose CNN-based solutions for retinal vessel segmentation with incorporated attention mechanism. Attention is often introduced through attention blocks that vary in design and position in a base architecture \cite{guo_2021_saunet, li_2021_fanet, zhang_2019_agnet, wang_2019_deunet}. 

According to \cite{guo_2022_attention} there are four main attention categories in computer vision: spatial, channel, temporal, and branch attention, with the first two being used in segmentation from fundus images \cite{guo_2021_saunet}. While spatial attention weights the importance of learned spatial features, channel attention weights the importance of in-channel image features. In \cite{guo_2021_saunet, li_2021_fanet} the authors propose using a single spatial attention block to selectively focus on important features learned by the encoder. To calculate attention maps, average and max pooling are used to derive intermediate maps from feature maps. Intermediate maps are then concatenated and convolved to obtain final attention maps that are used to weight encoder output before passing it to the decoder. In \cite{li_2021_fanet} spatial attention is used to build a dual-direction attention block that looks for inter-feature dependences in horizontal and vertical directions. The proposed attention block is implemented using average pooling and inserted in each decoder block of UNet architecture. Similarly, in \cite{zhang_2019_agnet} attention blocks precede each upsampling layer in the decoder. The attention is incorporated in a guided filter \cite{he_2012_guided} that uses high-level encoder feature maps to help filter out low-level decoder feature maps. Work in \cite{wang_2019_deunet} modifies UNet by adding two encoders, one to encode spatial, and the other to encode context information. The network uses channel attention implemented as self-attention to learn the importance of feature dependencies over channel dimension.

\section{Performance analysis}
\label{sec:numerical-results}

Here, we describe a baseline dataset for retinal vessel segmentation for which we collect and compare the performance of different models. We discuss metrics considered in the paper and analyze collected results.

\subsection{Dataset}
The DRIVE data is a baseline dataset in retinal vessel segmentation from fundus images. It consists of 40 color fundus images of size 768×584 pixels, collected in a screening program for diabetic retinopathy in the Netherlands. To acquire the images, a non-mydriatic 3CCD camera with of 45$\degree$ field of view was used. Images in the datasets belong to adults of age 25 to 90 with completely developed retinal vascularization that may be damaged as a consequence of diabetic retinopathy. For machine learning purposes, each fundus image was manually labeled to acquire binary segmentation masks delineating blood vessels from the background. Each mask is accompanied by a FoV mask, which is often used to limit methods of performance calculation solely to the retinal area. A dataset defines a training-test scheme, resulting in mostly consistent comparison protocols over different methods benchmarked on this dataset. Example data images can be seen in Fig. \ref{fig:fundus-probmap-segmask}.

\subsection{Metrics}
The performance of retinal vessel segmentation is commonly evaluated using standard, pixel-wise metrics for semantic segmentation. Metrics are based on a confusion matrix containing information on truly and falsely classified pixels relative to segmentation ground truth. These include true positives ($TP$), true negatives ($TN$), false positives ($FP$), and false negatives ($FN$). Evaluation metrics such as classification accuracy ($Acc$), sensitivity ($Sen$), specificity ($Spec$), and F1 score are calculated based on the confusion matrix as shown in Eq. \ref{eq:acc}, Eq. \ref{eq:sen}, Eq. \ref{eq:spec}, and Eq. \ref{eq:f1}, respectively. The area under the ROC curve (AUC) is another commonly reported metric calculated by plotting the ROC curve for the confusion matrix and different thresholds and measuring the area beneath the curve.

\begin{equation}\label{eq:acc}
    Acc = \frac{TP + TN}{TP + TN + FP + FN}
\end{equation}
\begin{equation}\label{eq:sen}
    Sen = \frac{TP}{TP + FN}
\end{equation}
\begin{equation}\label{eq:spec}
    Spec = \frac{TN}{TN + FP}
\end{equation}
\begin{equation}\label{eq:f1}
    F1 = \frac{2 \cdot TP}{2 \cdot TP + FP + FN}
\end{equation}

Retinal vessel segmentation datasets, including the DRIVE dataset, exhibit a high class imbalance. This is due to a significantly larger number of pixels labeled as background, compared to those pixels labeled as blood vessels. For the DRIVE dataset, the average percentage of blood vessel pixels is 8.7\% without the FoV mask and with a standard deviation of 1.07. The value is produced by calculating the average percentage of blood vessel pixels for each of the 40 images in a dataset and then averaging those values. Class imbalance affects the choice of performance metrics \cite{chicco_2020_advantages}. For example, when calculating accuracy, a trivial segmentation model classifying all pixels as a background would on average have 91.1\% accuracy, leaving a range of approximately 8.9\% for inter-model comparison. The most suitable metrics should account for the presence of disbalance to avoid the strong influence of the major class classification performance in the result. Some class imbalance-aware metrics include the F1 score and Matthews Correlation Coefficient (MCC). While F1 is calculated based on TP, FN, and FP, MCC takes into account all four confusion matrix categories proportionally to the number of samples in both classes.


\subsection{Numerical results}
\begin{figure}[!h]
\centering
\includegraphics[width=\linewidth]{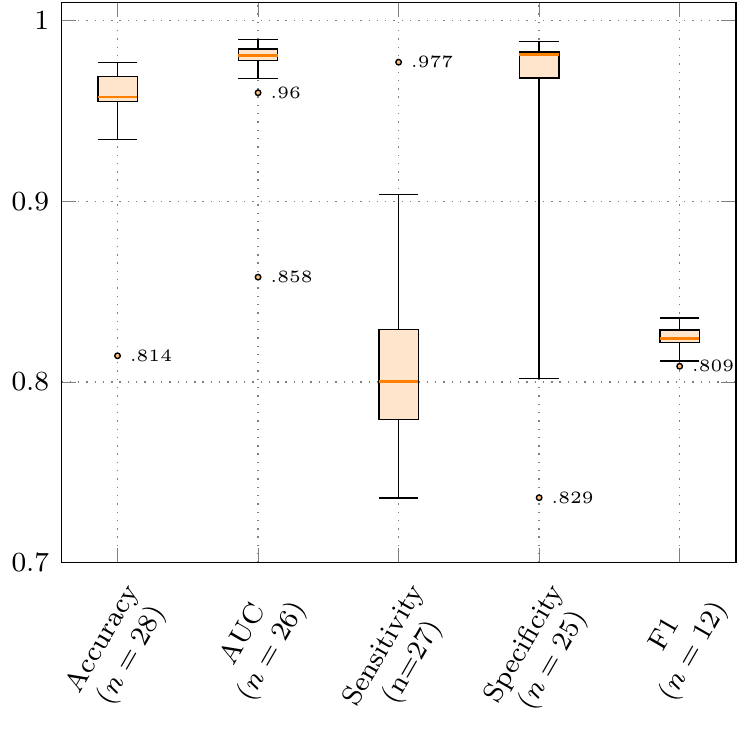}
\caption{Reported performance value distributions per metric with $n$ being the number of papers in which the particular metric is reported.\label{fig:metric-distributions}}
\end{figure}

In Fig. \ref{fig:metric-distributions} we show distributions of results reported in 39 papers on retinal vessel segmentation grouped by metric. All papers are published in the last three years in leading conferences and journals in the computer vision and machine learning fields. Some examples include IEEE Transactions on Medical Imaging (TMI), Conference on Computer Vision and Pattern Recognition (CVPR), and Conference on Medical Image Computing and Computer Assisted Intervention (MICCAI).


All distributions are presented as boxplots, or 5-point summary diagrams, with 5 characteristic points called lower whisker, lower quartile, median, upper quartile, and upper whisker. These points partition the whole distribution in the approximately 25\%, 50\%, and 75\% samples respectively. Since the lower and upper whiskers in Fig. \ref{fig:metric-distributions} do not correspond to the distribution minimum and maximum, outliers exist and are marked with dots. The exact values for all characteristic points can be seen in Table \ref{tabl:metric-distribution}, alongside other statistical properties, such as distribution mean value and standard deviation.

All metric values are in [0,1] range. According to the observed sample, the most reported metrics are classification accuracy (28/39), sensitivity (27/39), AUC (26/39), and specificity (25/39). All metrics except sensitivity have a comparatively low inter-quartile range of 1.5\% or below, meaning that 50\% of reported results differ by 1.5\% in metric value in case of accuracy or even less ($<$1\%) for other metrics. For accuracy, AUC, and F1 score, 99\% of distribution samples are clustered in the 5\% metric value range. While having a low inter-quartile range specificity has a more dispersed, right-skewed distribution with less than 25\% of methods achieving less than 90\% in background pixel classification. According to the comparison sample, sensitivity is the only metric having an almost symmetric and relatively dispersed distribution, not as tightly clustered around the median as is the case for accuracy, AUC, and F1. Dispersion of sensitivity distribution demonstrates differences in compared model abilities to correctly segment blood vessel pixels. 
Alongside here discussed metrics, other evaluation metrics reported in the papers include MCC ($n=2$), precision ($n=2$), and intersection over the union ($n=3$). However, due to the low number of samples, they are not considered in this paper.

In Table \ref{tabl:representative-paper-performance} we show results reported in the papers categorized into domain-knowledge, context-based, and attention-based approaches to gain preliminary insights into the efficiency of the approaches in retinal vessel segmentation. For each metric, we bolded out the best-reported result and underlined the second-best. Attention-based approaches yield the highest values in accuracy, AUC, and F1 score. The first three metrics take into account the classification efficiency of both background and blood vessel pixels and are not designed to account for class imbalance problems. As for the F1 score, attention-based approaches indicate superior results. However, it is not possible to derive reliable conclusions due to the small sample size. The second-best results in accuracy, AUC, and sensitivity are reported for domain-knowledge methods. Superior sensitivity results have been achieved in \cite{yeon_2019_vgn} using a combination of purely convolutional and graph convolutional networks. However, the same method reports the second-lowest result for specificity, introducing noise in the classification of background pixels. As sensitivity directly implicates the model's performance in blood vessel pixel segmentation, the analyzed sample indicates that incorporating domain knowledge into the network facilitates true positive classification.

\begin{table*}[!ht]
\centering
\begin{tabular}{c|c|c|c|c|c}
\toprule
& Accuracy & AUC & Sensitivity & Specificity & F1 \\
\midrule
$W_{lower}$     & 0.934 & 0.968 & 0.736 & 0.736 & 0.811 \\ 
$Q_{lower}$     & 0.955 & 0.978 & 0.779 & 0.977 & 0.822 \\
Median          & 0.958 & 0.980 & 0.800 & 0.981 & 0.824 \\
$Q_{upper}$     & 0.967 & 0.984 & 0.829 & 0.982 & 0.829 \\
$W_{upper}$     & 0.977 & 0.989 & 0.977 & 0.988 & 0.835 \\ 
IQR             & 0.014 & 0.006 & 0.050 & 0.006 & 0.007 \\
\hline\hline
$n$             &    28 &    26 &    27 &    25 &    12 \\ 
$\bar{x}$       & 0.955 & 0.971 & 0.804 & 0.957 & 0.824 \\
$\sigma$        & 0.029 & 0.034 & 0.050 & 0.064 & 0.007 \\
\bottomrule
\end{tabular}
\caption{
Numerical properties of metric distributions in Fig. \ref{fig:metric-distributions} with $n$ being a sample size, $\bar{x}$, and $\sigma$ mean value and standard deviation, respectively. The rest of the rows give numerical values for the box plot diagram. Median, lower whisker ($W_{lower}$), lower quartile ($Q_{lower}$), upper quartile ($Q_{upper}$), upper whisker ($W_{upper}$), and inter-quartile range (IQR). All floating values are rounded to 3 decimal digits.}
\label{tabl:metric-distribution}
\end{table*}

\begin{table*}[!ht]
\centering
\begin{tabular}{c|c|c|c|c|c|c}
\toprule
Method & Approach & Accuracy & AUC & Sensitivity & Specificity & F1 \\
\midrule
Cherukuri et al. \cite{cherukuri_2020_geometry} & Domain-knowledge & \underline{0.9723} & \underline{0.987} & \underline{0.8425} & 0.9849 & 0.822 \\
Shin et al. \cite{yeon_2019_vgn} & Domain-knowledge & 0.9271 & 0.9802 & \textbf{0.9382} & 0.9255 & - \\
Fan et al. \cite{fan_2019_imagematting} & Domain-knowledge & 0.96 & 0.858
& 0.736 & 0.981 & - \\
\hline
Wu et al. \cite{wu_2021_scsnet} & Context-based & 0.9697 & 0.9837 & 0.8289 & 0.9838 & - \\
Wang et al. \cite{wang_2019_contextaware} & Context-based & 0.8353 & - & - & 0.8419 & - \\
Wang et al. \cite{wang_2020_rvsegnet} & Context-based & 0.9681 & 0.9817 & 0.8107 & 0.9845 & - \\
\hline
Guo et al. \cite{guo_2021_saunet} & Attention-based & 0.9698 & 0.9864 & 0.8212 & 0.984 & \underline{0.8263} \\
Li et al. \cite{li_2021_fanet} & Attention-based & \textbf{0.9769} & \textbf{0.9895}
 & 0.8145 & \textbf{0.9883} & - \\
Zhang et al. \cite{zhang_2019_agnet} & Attention-based & 0.9692 & 0.9856 & 0.81 & 0.9848 & - \\
Wang et al. \cite{wang_2019_deunet} & Attention-based & 0.827 & 0.9567 & 0.794 & \underline{0.9772} & \textbf{0.9816} \\
\bottomrule
\end{tabular}
\caption{Performance metric values reported in representative papers for each of the approaches covered. Bolded values in each column are the highest reported, while the underlined ones are the second highest.}\label{tabl:representative-paper-performance}
\end{table*}

\section{Discussion}
\label{sec:discussion}

The most commonly used metrics in retinal vessel segmentation are accuracy, AUC, sensitivity, and specificity. Accuracy and AUC are class imbalance insensitive, making them suboptimal choices in retinal vessel segmentation, where images contain significantly more background than blood vessel pixels. Sensitivity and specificity measure the performance of a specific class and address the classification performance of blood vessels and background pixels. While sensitivity can be a significant indicator in classification, specificity is less informative as a result of class imbalance, and the tendency of the methods to classify the background with low-noise levels due to the larger training sample size for the negative class. While the F1 score is designed to address the class imbalance, the observed results on a limited sample number show low dispersion around the distribution mean, resulting in very similar results between compared solutions, indicating that class imbalance is too high to properly reflect on the F1 score. Alternative general-purpose metrics, such as MCC, or specialized metrics, such as connectivity proposed in \cite{li_2020_iternet}, might be more informative choices of model performance.

Considering the limited sample size for each of the retinal vessel approaches presented in this paper, domain-knowledge-based and attention-based approaches tend to yield better results compared to pure context-based approaches. While representative of domain-knowledge results utilizing graph convolutional neural networks achieves significantly higher sensitivity compared to other proposed solutions, it is computationally demanding to train and report lower specificity, generating also more noise in the background. With sufficient training resources, it could be possible to combine ideas from domain knowledge and attention-based approaches to maximize sensitivity while preserving high specificity.

\section{Conclusions}
\label{sec:conclusions}

In this study, we reviewed recent literature on deep learning methods for retinal vessel segmentation to identify the main conceptual improvements of encoder-decoder networks yielding efficient retinal vessel segmentation. We summarized how incorporating domain knowledge, richer segmentation context, and attention in deep neural networks can improve segmentation performance. To support the theoretical review, we provided a numerical analysis of the results and discussed approaches in the context of reported results. Preliminary results showed that among the isolated approaches pure context-based approaches are the least efficient, while domain knowledge and attention-based approaches have similar performance in general. However, certain ideas from those groups exhibit extraordinary numerical performance and thus could be combined in the future into a more efficient solution.

While this study identifies a subset of dominant conceptual approaches to retinal vessel segmentation, more approaches might be isolated from the literature, that were not discussed in this paper due to paper length constraints. Our assumption is that combining the advantages of different approaches, more efficient and robust solutions can be designed. 




\section*{Acknowledgment}
This paper has received funding from the European Union’s Horizon 2020 research and innovation programme under Grant Agreement number 85696. The paper has also been supported by the Ministry of Science, Technological Development, and Innovation of Republic of Serbia through project no. 451-03-47/2023-01/200156 “Innovative scientific and artistic research from the FTS domain”.

\bibliographystyle{IEEEtran}
\bibliography{references}


\end{document}